# A Digital Automatic Sliding Door with a Room Light Control System

Adamu Murtala Zungeru[1] and Paul Obafemi Abraham-Attah[2]

[1]School of Electrical and Electronic Engineering, University of Nottingham, Jalan Broga, 43500 Semenyih, Selangor Darul Ehsan, Malaysia
adamuzungeru@ieee.org

[2]Department of Electrical and Electronic Engineering, Federal University of Technology, Minna, Nigeria
pattah66@yahoo.com

## ABSTRACT

*Automatic door is an automated movable barrier installed in the entry of a room or building to restrict access, provide ease of opening a door or provide visual privacy. As a result of enhanced civilization and modernization, the human nature demands more comfort to his life. The man seeks ways to do things easily and which saves time. So thus, the automatic gates are one of the examples that human nature invent to bring comfort and ease in its daily life. To this end, we model and design an automatic sliding door with a room light control system to provide the mentioned needs. This was achieved by considering some factors such as economy, availability of components and research materials, efficiency, compatibility and portability and also durability in the design process. The performance of the system after test met design specifications. This system works on the principle of breaking an infrared beam of light, sensed by a photodiode. It consists of two transmitting infrared diodes and two receiving photo-diodes. The first one is for someone coming in and the second one is for someone going out of the room. The photodiodes are connected to comparators, which give a lower output when the beam is broken and high output when transmitting normally. The general operation of the work and performance is dependent on the presence of an intruder entering through the door and how close he/she is in closer to the door. The door is meant to open automatically but in a case where there is no power supply trying to force the door open would damage the mechanical control system of the unit. The overall work was implemented with a constructed work, tested working and perfectly functional.*

## 1. INTRODUCTION

Our contributions to the society are times fueled by personal experience complemented by knowledge of a particular field of study. Electronics system refines, extend or supplement human facilities and ability to observe, perceive, communicate, remember, calculate or reason. Electronic systems are classified as either analog or digital. Analog system change their signal output linearly with the input and can be represented on a scale by means of a pointer. On the other hand, digital instruments or circuits represent their output as two discrete levels ('1' or '0') and could show their output in a digital display either numerically or alphabetically.

The need for automation has come to stay and this date back to 1500 years when the first water pump for metal working rolling mills for coinage strips was developed [1] from then till date the automation world has continued to grow tremendously. Automation is the art of making processes or machines self-acting or self-moving, it also pertains to the technique of making a device, machine, process or procedure more fully automatic, it is a self-controlling or self-moving processes [2, 3].

Automation is usually characterized by two major principles: (1) mechanization, i.e. machines are self-regulated so as to meet predetermined requirements (a simple example of self-regulation can be found in the operation of a thermostatically controlled furnace); (2) continuous process, i.e. production facilities are linked together, thereby integrating several separate elements of the production process into a unified whole [4]. Automation in the electrical, electronics and computing world has grown rapidly of which it dates back to 1940 when the first electronics computing machine was developed [1]. This has aided humans as it basically reduces/eliminates human intervention, of which automatic light switching system also makes the list of automation in the electro-computing world.

Switch which is one part of this work may be the most ubiquitous mechanical devices in our technological society. Most every machine needs to be turned on or turned off at some point, and that's typically done by activating a switch. There are an incredible variety of switches. The most basic electrical switch completes or breaks a circuit depending upon what position it is in. Back in time we recall constructing science systems or experiments that required us to build a small electrical circuit that included a battery and a flashlight bulb. When the simple switch was moved, it completed the circuit and the bulb would glow. More complex switches work in the same basic manner. In addition to turning the machines on, they can change the speed of the motor or the strength of the lighting. Some switches work automatically, incorporating a tiny microprocessor that turns the machine on (or off) according to preset instructions. An example of a low-tech (pre-digital age) automatic switch is the thermostat used to maintain a set temperature in a home or building. These switches used a small glass vial containing a drop of mercury. When the temperature indicator moved beyond a certain level, the drop of mercury would move into contact with metal contact points that extended into the glass vial. Being a metal itself, the liquid mercury would complete the circuit and activate the furnace or air conditioner (as the case may be). Mercury switches are used rarely nowadays, and they should be disposed of using hazardous waste protocols due to the high toxicity of metallic mercury in the environment [5].

Another example of a switch that works automatically but a high tech one is an automatic room lighting system, which ranges from sound automatic room light system, infrared automatic room light system, temperature automatic room light system in addition to the basic principle stated above, the principle behind these automatic room light system is that the light turns ON and OFF automatically which is sensor dependent. For this system design, the principle behind this is that when a person enters a room, a light sensor placed at a particular location gets a pulse and the light comes ON and when the person goes out, the same sensor gets another pulse and the light goes OFF. The room light controller has a lot of domestic applications and besides, power is seriously conserved when using the unit since the light in the room is automatically switched off when nobody is in the room. The same opto-sensing stage is used to sense when someone enters or leaves the room. The sensors control the mono-stable multivibrator whose output activates the counter and comparator unit and switches ON the lights once there is any count, and OFF when the total count is zero (since the counter counts one place down for every person leaving the room) and one placed up when someone enters the room.

The other part of this work which is the automatic sliding door operates basically in conjunction with the automatic light switch of which it uses a motor for the sliding. The sliding of the door occurs when the same sensors placed at a particular location of the door is broken. A motor is incorporated to drive the system on breakage of a pulse at the receiving end. The mechanical arrangement of the door is done such that the door slides open with the control of D.C motor and metal bearing mechanism automatically upon detection of somebody approaching the door when the beam is broken. This system has both financial and security benefits. (1) Financial savings: By setting lights to come on only at certain times, you can reduce utility bills, and (2)

Security: With this kind of device, you can link lights to a timer so they come on when it gets dark. You'll never have to return home to a dark house.

One basic limitation of this device is that it encounters the problem that when more than one person enters the room one after the other the light sensors get more than one pulse and the light remains in the OFF state.

This research comprises both analog circuits and digital circuits. The opto sensing stage is an infrared transceiver at the base of the door which has a projected infra-red receiver on the opposite side of the door. The projected beam is broken by anybody entering the door. Once broken the infra-red receiver stage gives an output which triggers the two monostable multivibrators. Monostable (A) has a time constant of 20s and monostable (B), 10s. The output of both monostable goes to the input of an Exclusive-OR gate which allows only one relay to be active. Hence, enable timing of the motor in different directions, since opposite polarity voltage are furnished to the motor at different times, to open and close and door via a mechanical metal and bearing system. The application of this system cannot be overemphasized. Optical interruption, electronic timing functions and relay circuits, logic control (since the movement of the door is controlled logically and in particular directions). The system has both security application and luxury, since it is more comfortable and easy if the opening and closing of the door and switching of the room lights are done automatically. The system could be done and implemented in the building of school, hall, auditorium, banks, shopping malls and various departmental buildings.

## 2. RELATED WORK

The idea of using infrared signals to establish routes in communication networks between receivers and transmitters for the purpose of convenience, safety and guarantee of service is not new, but the application, cost, design method and reliability of the system varies. In [6], home automation systems as multiple agent Systems (MAS) were considered. In their work, home automation system was proposed that included home appliances and devices that are controlled and maintained in home management. Their major contribution to knowledge was to improve home automation, but not minding the cost of the entire system.

In a related work, [1, 2], in their paper also proposed an Internet Based Wireless Home Automation System for Multifunctional Devices. They proposed a low cost and flexible web-based solution but this system has some limitations such as the range and power failure.

In [2, 3], problems with the implementation of home automation systems were considered. Furthermore the possible solutions were devised through various network technologies. Several issues affecting home automation systems such as lack of robustness, compatibility issue and acceptability among older and disabled people were also discussed. Besides, much were treated in papers by Zungeru et al. [6, 7, 8] which consider the use of infrared rays to count the number of passengers in a car and also remotely control home appliances via short message services.

## 3. SYSTEM MODELLING AND DESIGN

This section will discuss the design procedure and the basic theory of components used for this work. The section is further divided into two sub-sections as component theory and system design and analysis.

### 3.1. Component Theory

In this sub-section, we describe and explain the theory behind the components used in this work ranging from their basic principles of operation to their application in this research work. The components used are: opto-devices, infrared emitters, photo diode, transistors, IC timers, comparators, memory latch and logic control, OR-gates, up/down counters, seven segment decoders, seven segment display, filaments, relays and other passive components.

### 3.1.1. Opto-Devices

Opto-devices convert light energy from one form to another. They are used for transmission of infrared rays, emission of light in different colors (i.e. LED's), sensing of light rays of different intensity (LDR's, photo-diodes and phototransistors), and for the conversion of light to different electrical quantities like current voltage and frequency. The various opto-devices are reviewed below.

### 3.1.1.1. Infrared Emitters

An infrared emitter just like the normal light emitting diode (LED) is generally a junction diode from the semi-conductor material, gallium arsenide phosphide. The Infrared action and type of rays is dependent on the type of semi-conductor doping used.

The infra-red type when furnished with appropriate voltage and current (which could be gotten from data sheets), emits infrared rays at a given wavelength. Typically the %mm LED emits infra-red of about 150mA at a voltage of about 1.7V D.C forward current. The symbol for the infra-red emitter is shown in Fig.1.

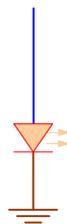

Fig. 1: infra-red emitter

### 3.1.1.2. Photo-diode

The photo diode consists of normal P-N junctions with a transparent window through which light can enter. A photo-diode is usually operated in reverse bias and leakage current increases in proportion to the amount of light falling on the junction. This effect is due to the semi-conductor and producing electrons and holes. Photo-diodes find application in counter circuit, scanners for disc, remote controls receivers' etc., the schematic symbol is shown in Fig 2.

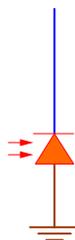

Fig. 2: schematic symbol photo-diode

### 3.1.1.3. Transistors

Transistors are active components used basically as amplifiers and switches. The two main types of transistors are. The bipolar transistors whose operations depend on the flow of both minority and majority carriers, and the unipolar of field effect transistors (called FETs) in which current is due to majority carriers only (either electrons or holes). The transistor as a switch operates in a Class A mode. In this mode of bias the circuit is designed such that current flows without any signal present. The value of the bias current either increased or decreased about its mean value by input signals (if operated as an amplifier), or ON and OFF by the input signal if operated as a switch Fig. 3 shows the transistor as a switch.

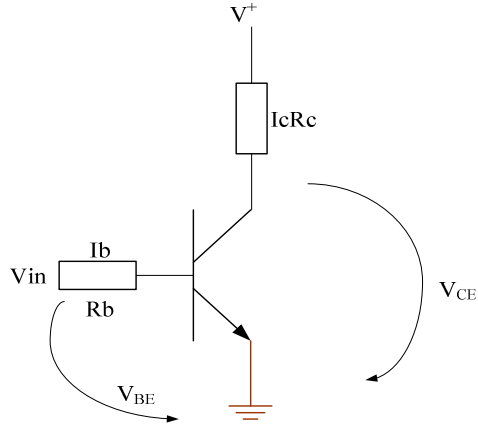

Fig. 3: transistor as a switch

For the transistor configuration, since the transistor is biased to saturation

$V_{CE} = 0$, when the transistor is ON,

This implies that,

$$V+ = I_C R_C + V_{CE} \tag{1}$$

$$V_{in} = I_B R_B + V_{BE} \tag{2}$$

$$hfe = \frac{I_c}{I_b} \tag{3}$$

$$R_b = \frac{V_{in} - V_{BE}}{I_b} \tag{4}$$

Where $I_c$ = collector current, $I_b$ = base current, $V_{in}$ = input voltage, $V^+$ = Supply voltage, $V_{CE}$ = collector emitter voltage, $H_{fe}$ = current gain and $V_{BE}$ = Base emitter voltage.

## 3.2. System Analysis and Design

### 3.2.1. Principle of Operation

This system works on the principle of breaking an infrared beam of light, sensed by a photodiode. It consists of two transmitting infrared diodes and two receiving photo-diodes. The first one is for someone coming in and the second one is for someone going out of the room. The photodiodes are connected to comparators, which give a low-output when the beam is broken and high output when transmitting normally. The low output of the comparator is used to trigger a mono-stable multi-vibrator, which is connected to a D-flip-flop, AND gates, OR gates, up/down counter and the display. The light would come ON, if the output of the up/down counter is not equal to zero and OFF, if the output of the up/down counter gets back to zero, as the last person leaves the room. The output of the comparators, also triggers two timers simultaneously, but both have different time constant. The first timer has a time constant of 5S while the second timer has a time constant of 10Secs. When the timers are triggered by breaking the beam, the first timer drives the transistor switch that controls the opening of the door. Hence, the door opens in 5 Seconds. The output of the timer is fed to an exclusive OR gate which gives an output only when the inputs are not same logic. When the timers are triggered the input into the EX-OR gate is the same logic, but after the first 5Seconds, the output of the first timer is low, and the output of the second is still high because it has a time constant of 10 seconds, so the EX-OR gate switches another relay that closes the door for 5 seconds. The door is closed by reversing the polarity of the supply to the motor. Hence the same sensors are used to switch the lights ON and OFF, open and close the door automatically

### 3.2.2. Power Supply Stage

All stages in the system section use 12V D.C. The power supply stage is a linear power supply type and involves a step down transformer, filter capacitor and voltage regulator

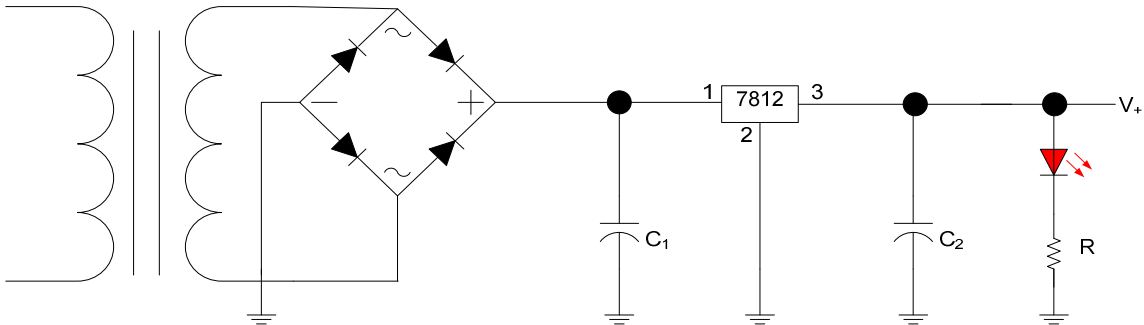

Fig. 4: Power supply stage.

For the circuit we would require a 7812 voltage regulator as shown in the figure above which gives the required output of +12V. The voltage regulator regulates above its required output voltage, if the voltage is below, its required output voltage would be passed out without being regulated. For example for a 7812, if the unregulated input voltage is greater than 12V, it will be regulated to 12V, but if it's less than 12V, for example 9V, the 9V unregulated will be outputted.

### 3.2.2.1. Design Analysis

If the unregulated input of the 7812 is greater than the required output by a factor of 4, that is 12+4=16, the voltage regulator IC, starts getting hot and will be damaged. Hence we will need an input into the 7812 to be approximately 16V.

Since, the diode drops 0.6V and we have 4 rectifying diodes forming the full wave bridge, the voltage drop will then be $0.6 \times 4 = 2.4V$
For a peak voltage of 16+2.4=18.4V peak.
For the r.m.s voltage $= \dfrac{18.4}{\sqrt{2}} = \dfrac{18.4}{1.4} = 13.143V$

Hence a transformer of a preferred value of 15V was employed, i.e. 220V/15V transformer
Assuming a ripple voltage of 15%

$$dv = \dfrac{15}{100} x18.4V \quad = \quad 2.76$$

$$dt = \dfrac{1}{2f} = \dfrac{1}{100} = 0.01$$

$$C_1 = C_1 = \dfrac{1 \times 0.01}{2.76} = 3.623 \times 10^{-3} F = 3300 \mu F$$

A preferred value of 3300μF was however employed.

To reduce the ripple left, a compensating capacitor $C_2$ was used and a 4065μF was employed

### 3.2.3. Sensor Stage

### 3.2.3.1. Transmitter Stage

This stage is not too complex since no coding is involved. Basically it is made up of infrared emitters and passive component like a resistor. The infrared diode is forward biased to meet the electrical conditions, which it operates. Fig below shows the infrared emitter stage.

$V^+$ = Power supply = +9V, $R_1$ = Limiting resistor

For the diode to be forward biased, the maximum forward voltage $V_F$ = 1.7V

Maximum forward current, $I_F$ = 150mA

The resistor $R_1$ would therefore be

$$R_1 = \frac{(V)-(V_F)}{I_F} = \frac{9-1.7}{150mA}$$

$R_1 = 48.67 \Omega$

A preferred value of $33 \Omega$ was used

### 3.2.3.2. Receiver Stage

This stage is made up of 2 comparators, 2 photo-diodes and 1 Infrared. The photo-diode reduces in resistance when it senses light and vice-versa. This principle is used for sensing stage which means that if the photo diode is connected in a voltage divider mode, and a resistance of the photo-diode reduces when it senses light and vice-versa, it means the voltage increases when it does not sense light. This change in voltage can serve as an input to a voltage comparator which is a desired output. The light source for the photodiode in this case is an infrared emitting diode. The comparator functions in the following manner. It gives a high output when the voltage in a non-inverting input is greater than the voltage at the inverting input and it gives a low output when the voltage at the inverting input is greater than or equal to the voltage in the non-inverting input so a photo-diode can be connected to the comparator as shown in Fig. 6 below.

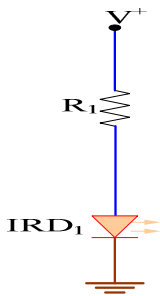 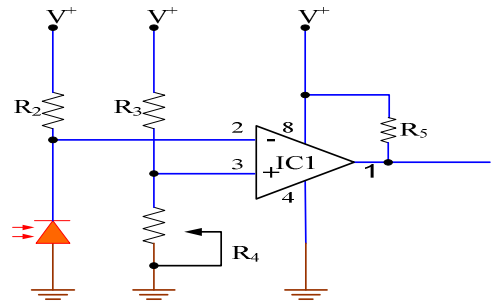

Fig. 5: Transmitter stage                    Fig. 6: photodiode connected to a comparator

$R_3$ is set to $1M\Omega$
$V^+$ = 12V

The voltage drop across photodiode = $\frac{(R\ photodiode) \times (V)}{R\ photodiode + (R\ photodiode)}$

$V_{photodiode} = \frac{(1M) \times (12)}{(1M)+(1M)}$

Let $V_3$ = 4V to serve as a reference voltage to pin 3 which is lower than the voltage at pin 2

$R_4 = 1k$

$$V_3 = \frac{(R) \times (V)}{(R) + (R)}$$

4K + 4R = 12R, 4K = 12R − 4R = 8R and R = 4/8 = 1/2K = 500 $\Omega$

A variable resistor of 4.7K was employed so that the maximum distance at which the photodiode can sense the infra-red can be achieved.

NB. The 1k resistor is the Pull Up resistor as it is required from the data book of the manufacturer.

### 3.2.4. Sensor Control Stage

### 3.2.4.1. Mono-Stable Stage

The mono-stable stage generates one shot of clocking pulse each time the sensor detects somebody entering the room. The mono-stable is triggered from the output of a comparator which senses the breaking of the beam and sends its clock to the input of a counter. Fig. 7 below shows a mono-stable.

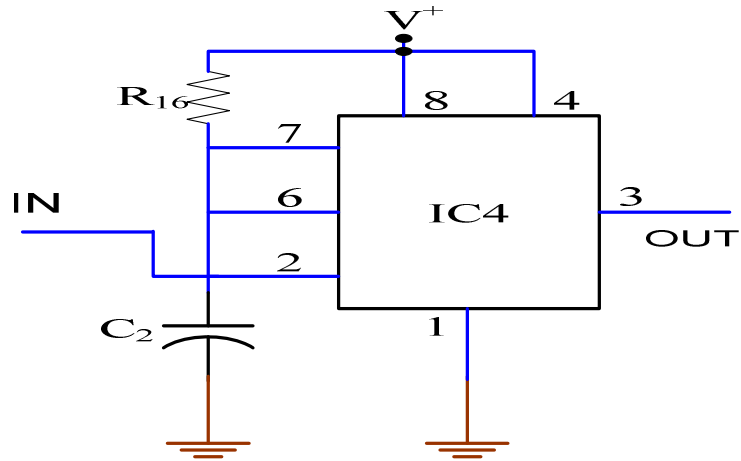

Fig. 7: Mono-Stable.

Since the formula to calculate the time for mono-stable is T = 1.1RC, and the time duration of the mono-stable is 10 Second (To allow for fast triggering of the counter). Assuming C = 100µF, Then T = 10 Second.

$$R_{16} = \frac{10}{1.1 \times 100 \times 10^{-6}} = \frac{10 \times 10^6}{110} = 90909.1 \, \Omega = 91 K\Omega$$

A preferred value of 100K$\Omega$ was used for R. To deactivate the second sensor, the 10 seconds mono-stable serves as an input to an OR gate, with the output of the sensor, so that as the mono-stable is timing irrespective of the output of the sensors, the output of the OR gate is always high, since the output of the mono-stables output is high, then the second sensor is inactive for the ten seconds allowing someone to pass without affecting the count of the other up/down counter. The Fig. 8 shows the circuit diagram.

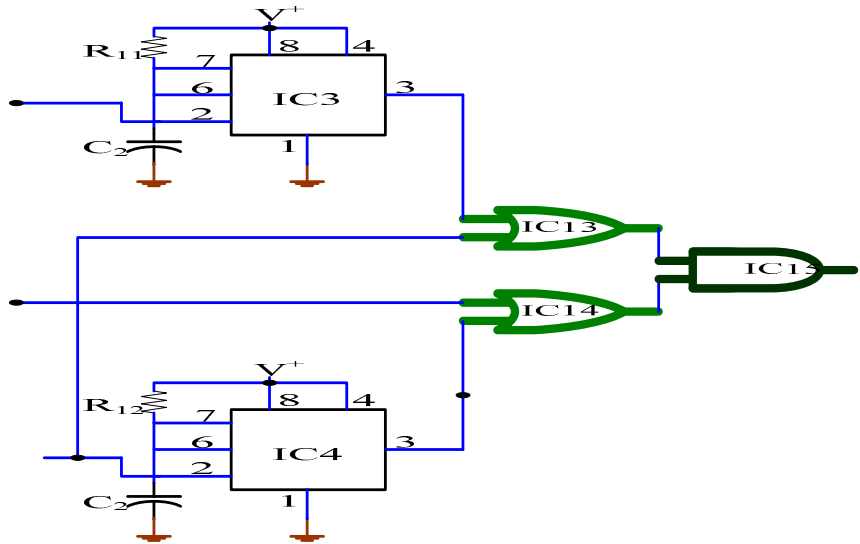

Fig. 8: circuit diagram described above

### 3.2.5. Counter Stage/ Display Unit

The counter employed is the 74190 BCD DECADE COUNTER which has two important input; Down/UP pin and the clock pulse pin. The 74190 counts up when it receives a clock pulse and the Down/up pin is low and count down when the down/Up is high. So to cause the counter to count the 10seconds, mono-stable triggered by the exit sensor is connected to the Down/up pin, so that when someone breaks the infrared beam while exiting, the Down/up pin will be high, causing the counter to count down. Similarly, when someone breaks the entry sensor, while entering the room, the counter will count up, since the Down/up counter is low. The output of the OR-gate, serves as an input to an AND-gate. This is to clock the counter. This counter counts to a maximum of 99, which means the maximum number of people in the room at a time should not exceed 99. Else the counter will not count properly, or will lose count of those in the room. The 74190 are cascaded to give a maximum count of 99, by connecting the ripple carry pin to the clock on the next counter. The 74190 counter outputs binary coded, this has to be converted to 7-segment format, by the 7 segment decoder driver, the 7447, 7 segment display, displays the count. The 7-segment display is connected with the common anode mode. This is because the output of the 7447 is active low. Fig 9 shows the diagram below

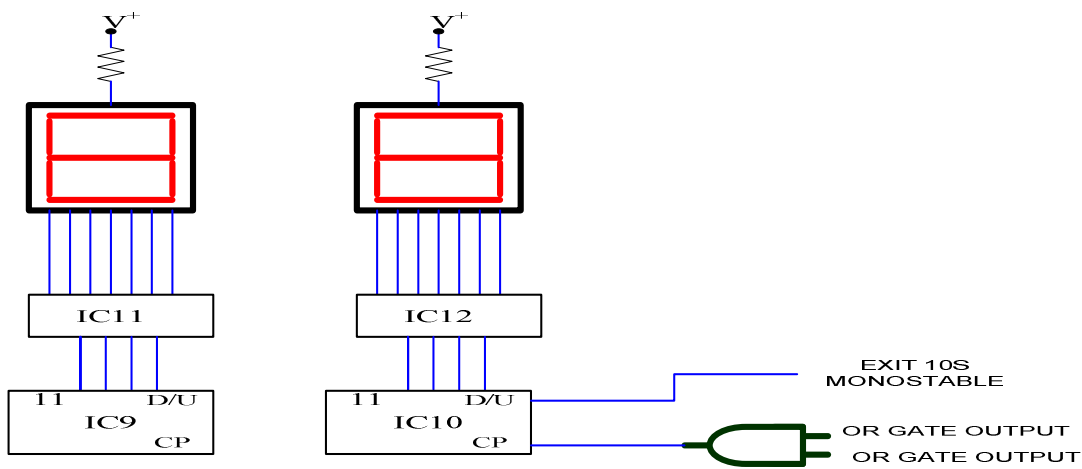

Fig. 9: circuit diagram of the counter stage/display unit

### 3.2.6. Driver Control Stage

This consists of a mono-stable multivibrator used to create a time constant which allows opening and closing of the door, and a combinational logic circuit which is used to determine when the door should lock or open. Two monostable stages are employed. The first opens the door, while the second closes the door, the stages are as shown.

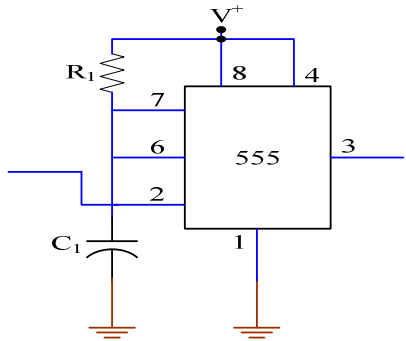 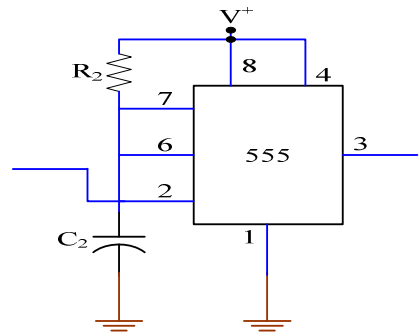

Fig. 10(a): Monostable 1                         Fig. 10(b): Monostable 2

The formula for the time constant is given as
T = 1.1RC
Where T = time constant, $R_1$ and $R_2$ = the resistors in the timing circuit, $C_1$ and $C_2$ = Timing capacitors. Both mono-stable stages are triggered simultaneously; mono-stable 1 has a time of 5seconds. By choosing an appropriate value of C, say: C = 100µF
Therefore, T = 1.1R × 100µF

For T = 5s, R = $\dfrac{5 \times 10^6}{110}$, R = 0.045 $\times 10^6$ Ω = 45KΩ

Preferred value of 47KΩ was used. The 5s constant was fixed because it is estimated that it would take approximately 5s to walk through the door from the sensor in a worst case condition. The design is such that exclusive gates give an output when the inputs are un-identical, making the inputs unequal is achieved by having longer time of mono-stable 2 hence durations of 10s was chosen so that the remaining 5s would switch another relay which closes the door.

For mono-stable 2, assuming C = 100µF, and we set T = 10s

R = $\dfrac{10}{1.1 \times 100 \mu F}$ = 90.9KΩ

A preferred value of 100kΩ was used.
The 100K preset was used, though there might be a little time difference, but it is negligible.
The exclusive OR gate determines when the door closes or opens, When the first timer mono-stable 1 output is high, the second is high too, so the ex-OR gives a low output after 5s, the output of mono-stable 1, goes low and the inputs to the ex-OR becomes un-identical and it gives a high output, the ex-OR gate is shown below;

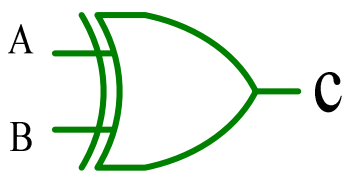

| A | B | C |
|---|---|---|
| 0 | 0 | 0 |
| 0 | 1 | 1 |
| 1 | 0 | 1 |
| 1 | 1 | 0 |

Fig. 11: Logic diagram and Truth Table of an Ex-OR gate

### 3.2.7. Driver Stage

The driver stage consists of a switching transistor stage which switches the relays that control the DC motor. Two switching transistors are used. This is to enable alternate switching to allow for movement of the motor in dual directions. The circuit diagram of the switching transistor is shown below in Fig. 12.

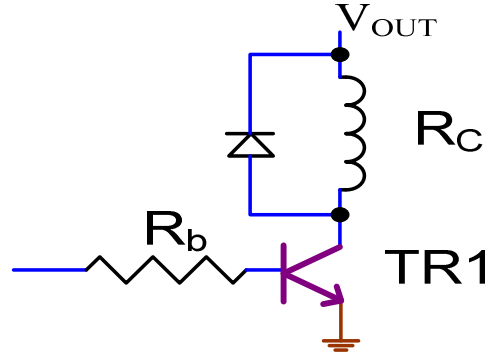

Fig. 12: Circuit diagram of a transistor as a switch

Where, $R_C$ is the coil resistance of the relay, $R_B$ the base resistor, $V^+$ is supply voltage = 12V, $V_{in}$ is input voltage = 10V and TR1 is switching transistor.

From $V^+ = (I_C \times R_C) + V_{CE}$, since the transistor is acting as a switch, which implies that $V_{CE} = 0$

$Ic = \dfrac{V^+}{Rc}$, and Since $R_C = 400\Omega$, it implies that $Ic = \dfrac{12}{400} = 0.03A$

$h_{FE} = \dfrac{I_c}{I_B}$, and $I_B = \dfrac{I_C}{h_{FE}} = \dfrac{0.03}{200} = 100\mu A$ ($h_{FE} = 200$ from data sheet)

Also $V_{in} = I_B \times R_B + V_{BE}$, i.e. 10 = 100µF x $R_B$ + 0.6V (where 0.6 = $V_{BE}$ for silicon)

Therefore $R_B = \dfrac{10 - 0.6}{100\mu} = 94000\Omega$. Therefore, $R_B = 94K\Omega$

Preferred value of 100K$\Omega$ was employed

The diode across the coil is used to protect the transistor against back E.M.F which might arise from coil since it is an inductive load

### 3.2.8. Driver Control Stage

This comprises of the transistor to switch the relay, an OR gate is used to combine the counter output and give an output, that is if any of the outputs of the counter is high, the output of the OR-gate is high, this is used to switch a relay that puts on the light automatically, so the light only goes off when the counter counts zero. The truth table is shown in Table 1.

In this case RC which is the collector resistance is the resistance of the relay coil which is 400 $\Omega$ for the relay type used in this system. Hence $R_C = 400\Omega$, $V^+ = 12V$ (Regulated voltage from power supply). $V_{BE} = 0.6V$ (Silicon transistor), $V_{CE} = 0V$ when the transistor is switched, $V_{in} = 10.3V$ (From the timer output), and $h_{FE} = 350$.

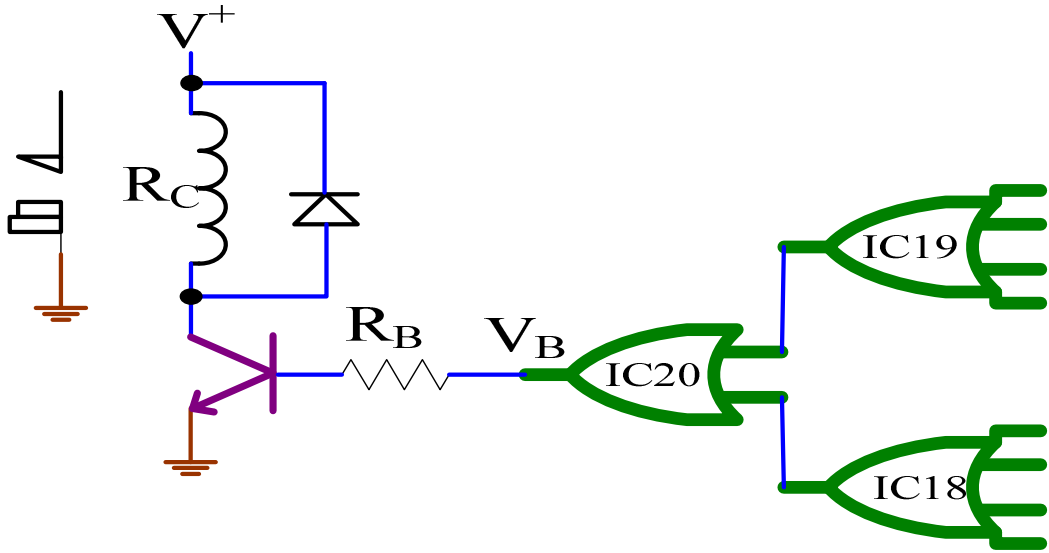

Fig. 13: Circuit diagram of the driver control stage.

From the transistor equations $V_+ = I_C R_C + V_{CE}$

To contain the transistor to cutoff region, it implies $V_{CE} = 0$, $V_+ = I_C R_C$, $I_C = V_+/R_C = 12V/400\Omega$ = 0.03A. But $h_{FE} = \dfrac{I_c}{I_b}$, and $I_b = \dfrac{I_c}{h_{FE}}$. The transistor used is the BC 337, with $h_{FE}$ of 350

Therefore, $I_b = \dfrac{0.03}{350} = 8.57 \times 10^{-5}$ A

$V_B = 12$ V, and $V_{BE} = 0.6$V (for silicon transistor) i.e. $12 = (8.57 \times 10^{-5}) R_B + 0.6$

Therefore, $R_b = \dfrac{12 - 0.6}{8.57 \times 10^{-5}} = 133022.17\,\Omega = 133\,K\Omega$

A preferred value of 130K$\Omega$ was employed

Truth Table 1

| Count | | | | Display | Light |
|---|---|---|---|---|---|
| A | B | C | D | | |
| 0 | 0 | 0 | 0 | 0 | OFF |
| 0 | 0 | 0 | 1 | 1 | ON |
| 0 | 0 | 1 | 0 | 2 | ON |
| 0 | 0 | 1 | 1 | 3 | ON |
| 0 | 1 | 0 | 0 | 4 | ON |
| 0 | 1 | 0 | 1 | 5 | ON |
| 0 | 1 | 1 | 0 | 6 | ON |
| 0 | 1 | 1 | 1 | 7 | ON |
| 1 | 0 | 0 | 0 | 8 | ON |
| 1 | 0 | 0 | 1 | 9 | ON |

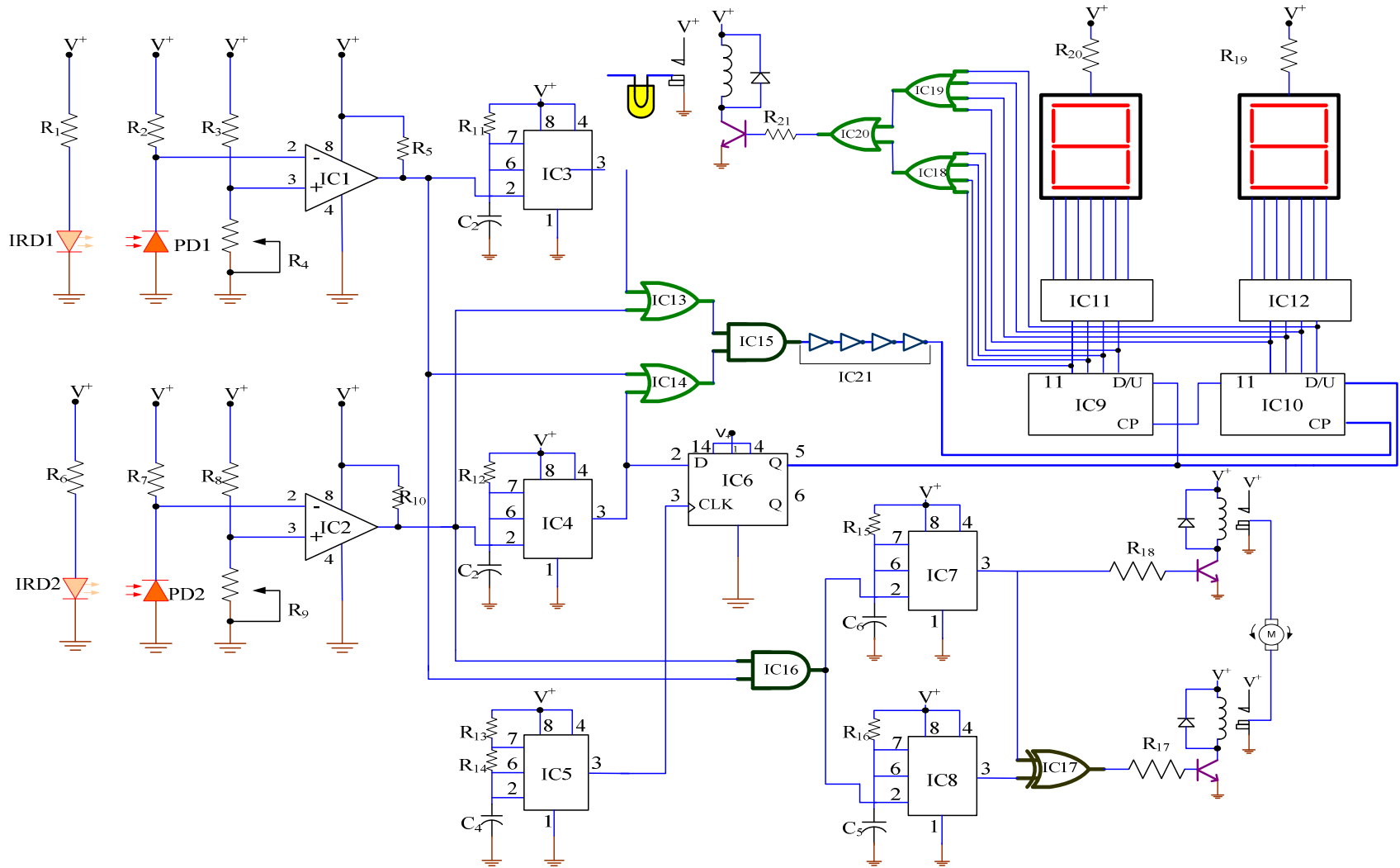

**Fig. 14: Complete Circuit Diagram of the Digital Automatic Sliding Door with Room Light Control System**

# 4. SYSTEM CONSTRUCTION, TESTING AND DISCUSSIONS

## 4.1. Construction and Layout

The construction of this system is in 2 stages, the soldering of the components and the coupling of the entire system to the casing. The power supply stage was first soldered, and then the transmitter and receiver stage and all the other stages were soldered. The circuit was soldered in a number of patterns that is, stage by stage. Each stage was tested using the multi-meter to make sure it is working properly before the next stage is done. This helps to detect mistakes and faults easily. The soldering of the circuit was done on a 10cm by 24cm Vero-board. The second stage of the system construction is the casing of the soldered circuit. This system was cased in a transparent plastic glass, this makes the system look attractive, and it helps in marketing the project because the circuit has to be attractive before someone would want to know what it does. The casing has special perforation to ensure the system is not overheating, and this will aid the life span of the circuit. Fig. 15 shows the location of the components on the Vero board.

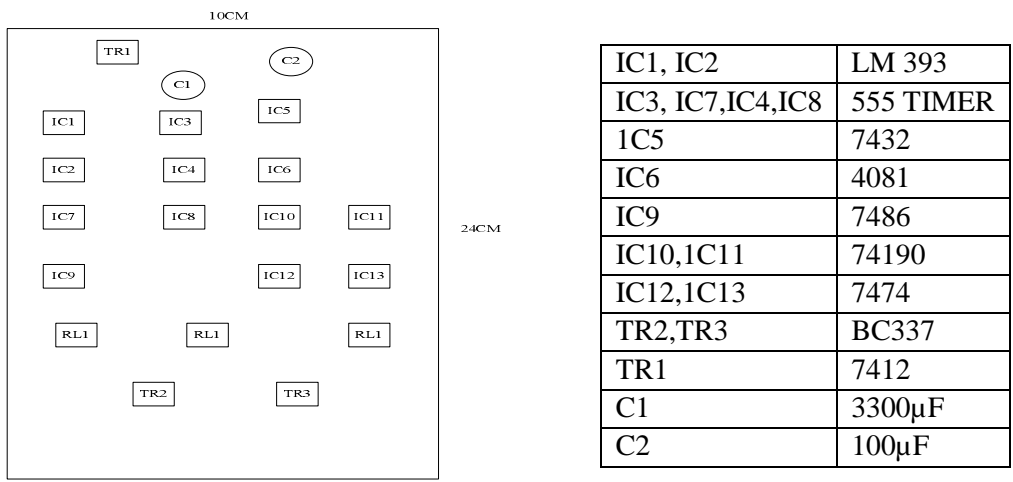

| IC1, IC2 | LM 393 |
|---|---|
| IC3, IC7,IC4,IC8 | 555 TIMER |
| 1C5 | 7432 |
| IC6 | 4081 |
| IC9 | 7486 |
| IC10,1C11 | 74190 |
| IC12,1C13 | 7474 |
| TR2,TR3 | BC337 |
| TR1 | 7412 |
| C1 | 3300µF |
| C2 | 100µF |

Fig. 15: Location of the components on the Vero-board

## 4.2. Testing

The physical realization of the system is very vital. This is where the fantasy of the whole idea meets reality. The designer will see his or her work not just on paper but also as a finished hardware. After carrying out all the paper design and analysis, the project was implemented and tested to ensure its working ability, and was finally constructed to meet desired specifications. The process of testing and implementation involved the use of some equipments such as digital multi-meter, signal generators and Oscilloscope. The digital multi-meter basically measure voltages, resistance, continuity, current, frequency, temperature, and transistor $h_{FE}$. The process of implementation of the design on the board required the measurement of parameters like, voltages, continuity, resistance values of the components and in some cases frequency measurement. The digital multi-meter was used to check the various voltage drops at all stages in the system, and most importantly the infra-red receiver stage, to help check the references in the comparator circuit. Also the digital multi-meter was used for troubleshooting the soldering and coupling. The oscilloscope was used to monitor the behaviors of the oscillators and also in checking the accuracy of the input and output voltages at each stage. In some cases, we used the signal generators to test the flow of signals before inputting the real signals generated from part of the system designed. We also use the oscilloscope to check for the results generated by the oscillators. The Transmitter pulses the infra-red (LED) with a waveform from with the output voltage volt of the 555 Timer. This results in a transition from a low to high state.

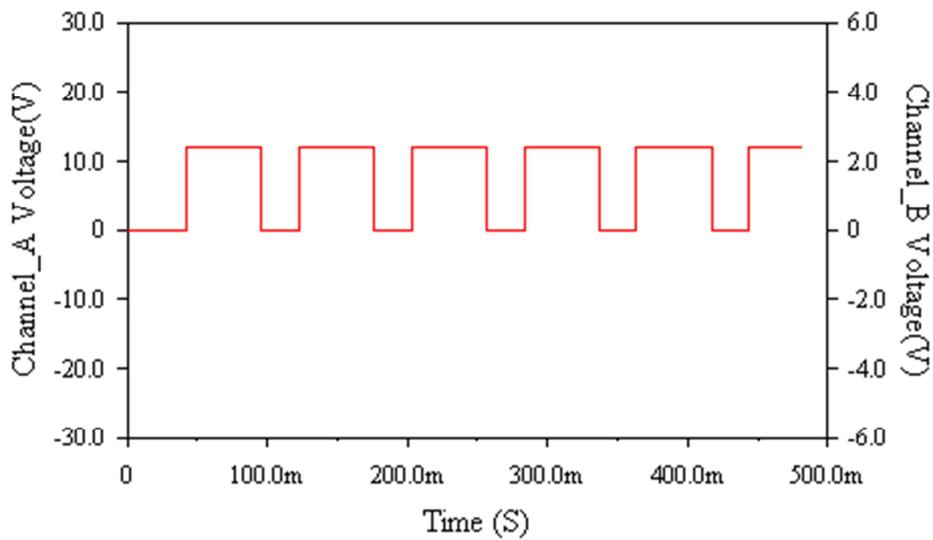

Fig. 16: Waveform generated

### 4.3. Implementation

The implementation of this system was done on the breadboard. The power supply was first derived from the bench power supply in the university electronics laboratory to confirm the workability of the circuit before the power supply stage was soldered. Stage by stage testing was done according to the block representation on the breadboard, before soldering of circuit commenced on the Vero board. The various circuits and stages were soldered in tandem to meet desired workability of the system. This system was coupled using plastic glass. The casing material being plastic glass is designed with special perforation and vents to ensure the system is not overheating and to give ecstatic value. The transceiver part of the system excluding the mechanical part is as shown in Fig. 17 below. The electronic circuit on the Vero board was securely screwed to the inside base of the case. Finally, the power switch, output LEDs, transmitter and receiver were put in their slots. This is what the finished work looks like:-

**Top**

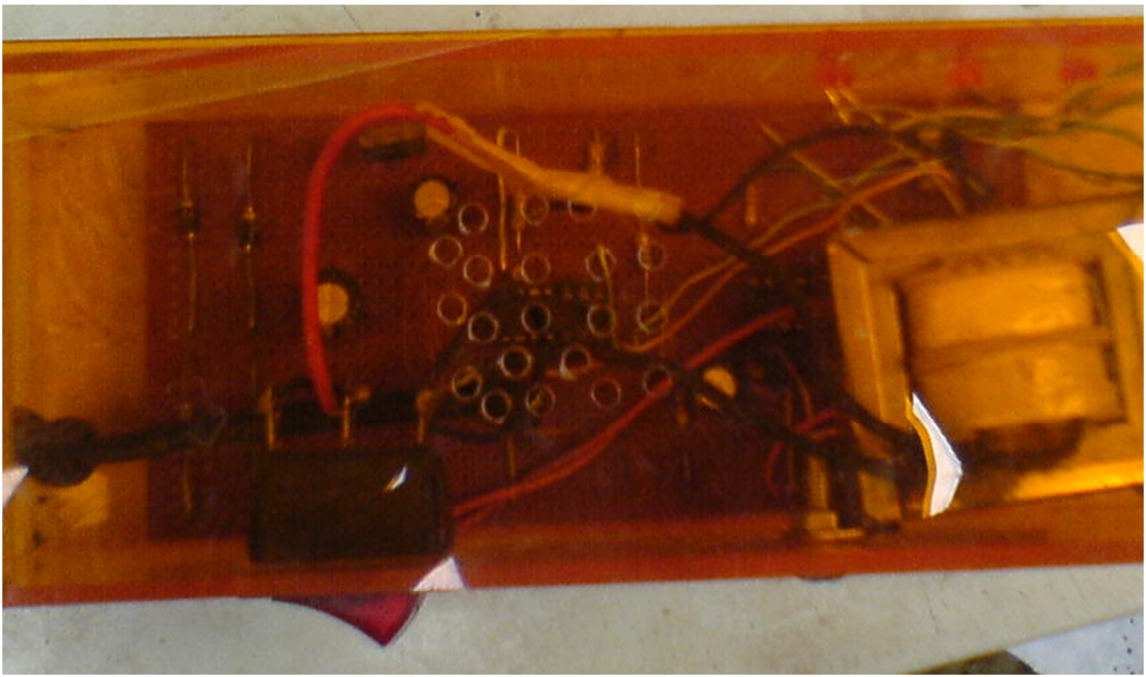

**Isometric view**

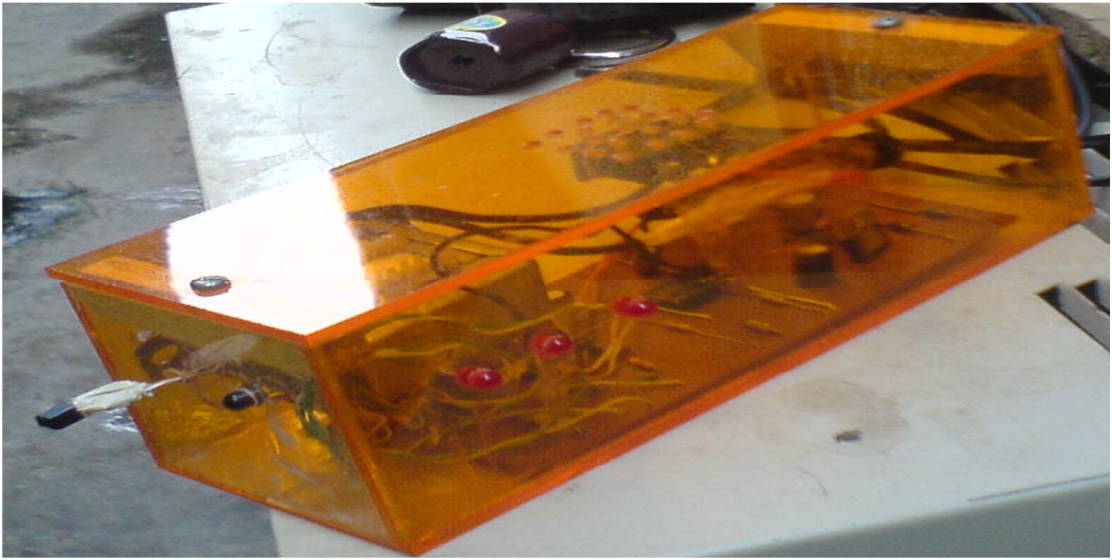

Fig. 17: Real-time implementation of the system

## 4.4. Problem Encountered

Several problems were encountered during the project. The problems range from design problems to implementation problems and also construction problems. The major problems are as follows: (1) Inability to turn the motor in both directions. This was the first design challenge the system posed. The problem was solved by using two relay drivers and combinational logic circuit (i.e. the exclusive OR gate). The relay contacts are arranged in such a way that their polarity is reversed when alternative signals occurs. (2) The relays were irked at some instance. This was discovered to be due to hysteresis. The problem was solved by using filter capacitors at the output of the comparator and across the relays. (3) Also, exact calculated values were used instead and this caused drifts in time constant of the timers but these drifts were negligible since they were within operating range and didn't disturb the opening and closing of the door. (4) The counter IC's, 74190, generated noise which affected the count sequence. This was solved by proper filtering of the outputs using capacitors. Other problems include soldering and measurement errors but these problems were solved by proper troubleshooting serious care in the construction of the project.

## 5. CONCLUSIONS

The system which is the design and construction of an automatic sliding door was designed considering some factors such as economy, availability of components and research materials, efficiency, compatibility, portability and also durability. The performance of the system after test met design specifications. The general operation of the system and performance is dependent on the presence of the person entering through the door and how closer he/she is to the door. The door is meant to open automatically but in a case where there is no power supply trying to force the door open would damage the mechanical control system of the unit. Also the operation is dependent on how well the soldering is done, and the positioning of the components on the Vero board. The IC's were soldered away from the power supply stage to prevent heat radiation which, might occur and affect the performance of the entire system. The construction was done in such a way that it makes maintenance and repairs an easy task and affordable for the user should there be any system breakdown. All components were soldered on one Vero-board which makes troubleshooting easier. The design of the automatic slide door involves;

research in both analog and digital electronics. Intensive work was done on timers and logic control circuits. Also research was done with relays and Opto-devices (e.g. Photodiodes, photo cells etc.). In general, the system was designed, and the real time implementation was done with a photo-type of the model.

## ACKNOWLEDGEMENTS

The author would like to thank Col. Muhammed Sani Bello (RTD), OON, Vice Chairman of MTN Nigeria Communications Limited for supporting the research.

**Authors**

**Engr. Adamu Murtala Zungeru** received his BEng in electrical and computer engineering from the Federal University of Technology (FUT) Minna, Nigeria in 2004, and his MSc in electronic and telecommunication engineering from the Ahmadu Bello University (ABU) Zaria, Nigeria in 2009. He is currently a lecturer two (LII) at the FUT Minna, Nigeria, a position which he started in 2005. He is a registered engineer with the Council for the Regulation of Engineering in Nigeria (COREN), a professional member of the Institute of Electrical and Electronics Engineers (IEEE), and a professional member of the Association for Computing Machinery (ACM). He is currently a PhD candidate in the department of electrical and electronic engineering at the University of Nottingham, Malaysia Campus.

His research interests are in the fields of swarm intelligence, routing algorithms, wireless sensor networks, energy harvesting, automation, and energy management for micro-electronics.

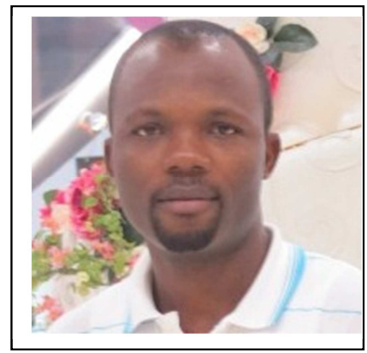